\documentclass[letterpaper, conference]{IEEEtran}
\makeatletter
\def\ps@headings{%
\def\@oddhead{\mbox{}\scriptsize\rightmark \hfil \thepage}%
\def\@evenhead{\scriptsize\thepage \hfil \leftmark\mbox{}}%
\def\@oddfoot{}%
\def\@evenfoot{}}
\makeatother
\usepackage{cite}
\usepackage{csquotes}

\usepackage{outlines}

\ifCLASSINFOpdf
   \usepackage[pdftex]{graphicx}
   \graphicspath{{../pdf/}{../jpeg/}}
   \DeclareGraphicsExtensions{.pdf,.jpeg,.png}
\else
   \usepackage[dvips]{graphicx}
   \graphicspath{{../eps/}}
   \DeclareGraphicsExtensions{.eps}
\fi
\ifCLASSOPTIONcompsoc
  \usepackage[caption=false,font=normalsize,labelfont=sf,textfont=sf]{subfig}
  \usepackage{subcaption}
\else
  \usepackage[caption=false,font=footnotesize]{subfig}
\usepackage[acronym, nowarn]{glossaries}
\makeglossaries
\newacronym{manet}{\textsc{MANET}}{mobile ad hoc networks}
\newacronym{dtn}{\textsc{DTN}}{delay-tolerant networks}
\newacronym{kytum}{\textsc{KyTuM}}{KeY and TrUst Management}
\newacronym{sol}{\textsc{SoL}}{Sea of Lights}
\newacronym{apdu}{\textsc{APDU}}{Application Protocol Data Unit}
\newacronym{nfc}{\textsc{NFC}}{Near-Field Communication}
\newacronym{one}{\textsc{ONE}}{Opportunistic Network Simulator}
\newacronym{imei}{\textsc{IMEI}}{International Mobile Equipment Identity}

\newacronym{wot}{\textsc{WoT}}{web-of-trust}

\usepackage{todonotes}
\usepackage{tabularx}
\usepackage[flushleft]{threeparttable}
\usepackage{array,booktabs,ragged2e}

\usepackage{tikz}
\usetikzlibrary{calc, positioning, arrows, shapes, shadows, trees, arrows.meta, 
decorations.pathreplacing, hobby, backgrounds} 
\usepackage{adjustbox}
\usepackage{epstopdf}
\usepackage{multirow}
\usepackage{wasysym}
\usepackage{gensymb}
\usepackage{amssymb}
\usepackage{pifont}

\usepackage{dblfloatfix}

\usepackage{amsthm}
\usepackage[hyphens]{url}
\usepackage{hyperref}
\hypersetup{
    colorlinks = false,
    allcolors = {red}
}

\theoremstyle{definition}

\pgfdeclaredecoration{dashsoliddouble}{initial}{
  \state{initial}[width=\pgfdecoratedinputsegmentlength]{
    \pgfmathsetlengthmacro\lw{.3pt+.5\pgflinewidth}
    \begin{pgfscope}
      \pgfpathmoveto{\pgfpoint{0pt}{\lw}}%
      \pgfpathlineto{\pgfpoint{\pgfdecoratedinputsegmentlength}{\lw}}%
      \pgfmathtruncatemacro\dashnum{%
        round((\pgfdecoratedinputsegmentlength-3pt)/6pt)
      }
      \pgfmathsetmacro\dashscale{%
        \pgfdecoratedinputsegmentlength/(\dashnum*6pt + 3pt)
      }
      \pgfmathsetlengthmacro\dashunit{3pt*\dashscale}
      \pgfsetdash{{\dashunit}{\dashunit}}{0pt}
      \pgfusepath{stroke}
      \pgfsetdash{}{0pt}
      \pgfpathmoveto{\pgfpoint{0pt}{-\lw}}%
      \pgfpathlineto{\pgfpoint{\pgfdecoratedinputsegmentlength}{-\lw}}%
      \pgfusepath{stroke}
    \end{pgfscope}
  }
}  

\tikzset{
  -|-/.style={
    to path={
      (\tikztostart) -| ($(\tikztostart)!#1!(\tikztotarget)$) |- (\tikztotarget)
      \tikztonodes
    }
  },
  -|-/.default=0.5,
  |-|/.style={
    to path={
      (\tikztostart) |- ($(\tikztostart)!#1!(\tikztotarget)$) -| (\tikztotarget)
      \tikztonodes
    }
  },
  |-|/.default=0.5,
}

\newcommand\copyrighttext{%
  \footnotesize \copyright{} 2018 IEEE. Personal use of this material is permitted. Permission from IEEE must be obtained for all other uses, in any current or future media, including reprinting/republishing this material for advertising or promotional purposes, creating new collective works, for resale or redistribution to servers or lists, or reuse of any copyrighted component of this work in other works. The official version can be found at \url{http://dx.doi.org/10.1109/LCN.2018.8638102}}
  
\newcommand\copyrightnotice{%
\begin{tikzpicture}[remember picture,overlay]
\node[anchor=south,yshift=10pt] at (current page.south) {\fbox{\parbox{\dimexpr\textwidth-\fboxsep-\fboxrule\relax}{\copyrighttext}}};
\end{tikzpicture}%
}

\begin{document}

%
\title{Sea of Lights: Practical Device-to-Device Security Bootstrapping in the Dark}

\author{\IEEEauthorblockN{Flor~\'{A}lvarez,
        Max~Kolhagen,
        and~Matthias~Hollick} 
        
\IEEEauthorblockA{Secure Mobile Networking Lab, TU Darmstadt, Germany, 
\{falvarez, mkolhagen, mhollick\}@seemoo.tu-darmstadt.de}
}

\maketitle
\copyrightnotice

\begin{abstract}
Practical solutions to bootstrap security in today's information and communication systems critically depend on centralized services for authentication as well as key and trust management.
This is particularly true for mobile users. 
Identity providers such as Google or Facebook have active user bases of two billion each, and the subscriber number of mobile operators exceeds five billion unique users as of early 2018.
If these centralized services go completely `dark' due to natural or man made disasters, large scale blackouts, or country-wide censorship, the users are left without practical solutions to bootstrap security on their mobile devices.
Existing distributed solutions, for instance, the so-called web-of-trust are not sufficiently lightweight. Furthermore, they support neither cross-application on mobile devices nor strong protection of key material using hardware security modules.
We propose \gls{sol}, a practical lightweight scheme for bootstrapping device-to-device security wirelessly, thus, enabling secure distributed self-organized networks.
It is tailored to operate `in the dark' and provides strong protection of key material as well as an intuitive means to build a lightweight web-of-trust.
\gls{sol} is particularly well suited for local or urban operation in scenarios such as the coordination of emergency response, where it helps containing/limiting the spreading of misinformation.
As a proof of concept, we implement \gls{sol} in the Android platform and hence test its feasibility on real mobile devices. We further evaluate its key performance aspects using simulation.

\end{abstract}


%
\IEEEpeerreviewmaketitle

\section{Introduction}
\label{sec:introduction}
Users of mobile applications heavily rely on third party providers to offer basic security services.
In fact, practical solutions to bootstrap security in today's information and communication systems critically depend on such third party security services for authentication as well as key and trust management.
These systems are commonly designed in a centralized fashion and scale to billions of users---making them attractive for application developers/providers as well as end users, which do not want to burden themselves with alternate means of securing their apps.

If these centralized services go completely `dark' due to natural or man made disasters \cite{Nepal.2015}, large scale blackouts \cite{Irma.2017}, or country-wide censorship \cite{HongKong.2014}, the users are left without practical solutions to bootstrap security on their mobile devices. As a result mobile app usage is severely restricted to support users in such scenarios.

Distributed solutions to bootstrap security such as the \gls{wot} exist. However, existing approaches are still complex and require educated users \cite{PGPusability}. Moreover, most current \gls{wot} solutions are ill-suited for cross-application support on mobile devices and do not support strong protection of key material by means of hardware security modules.

In this work, we propose \textbf{S}ea \textbf{o}f \textbf{L}ights (\gls{sol})\footnote{The name is inspired by silent protests such as candlelight vigils, where light is spread among the candles of a large group of people, effectively forming a `sea of lights'.}, a practical lightweight scheme for bootstrapping device-to-device security and for wirelessly spreading it to enable secure distributed self-organized networks. 
\gls{sol} is designed to complement existing self-organized network solutions by providing a lightweight and agile solution for decentralized authentication, key ma-nagement, and trust management.

Any third party mobile app can utilize the services of \gls{sol}, which offers an interface to access its security services. The security configuration can be performed on per-app granularity. 
Public/private key pairs generated by a third party application (=\textit{sub-keys}) can be authenticated with \gls{sol}.
Hence, our framework registers the public key of a sub-key, signs it, issues a certificate and is further responsible for its distribution.\\
To this end, \gls{sol} comprises two layers. 
(1) The \emph{Trust Management Layer}, which manages all operations related with trust relations. It is in charge of bootstrapping on demand, and of creating and maintaining trust relations.
(2) The \emph{Key Management Layer}, which is responsible for generating key material and managing the access to these keys in a secure fashion. 
This layer cares for, e.g., choosing an appropriate key storage according to the hardware capabilities and the available secure elements on the mobile devices.

In summary, the contributions of this paper are the following:
\begin{itemize}
\item We introduce \gls{sol}, a cross-application framework for bootstrapping security in device-to-device communication settings.
\item As part of \gls{sol}, we design and implement an automatic and decentralized key and trust management solution for mobile devices. It adapts to the hardware capabilities of the host device and is able to utilize hardware security solutions to further improve the security of the underlying keys. 
\item We evaluate the performance of \gls{sol} on real devices to demonstrate and test its feasibility in practice and further assess its key performance features by 
means of simulation. 
\end{itemize}

This paper is organized as follows. In Section \ref{sec:relatedwork} we summarize related work. In Section \ref{sec:assumptions}, we concisely introduce our motivating scenario, the system model, and the adversary model. Section \ref{sec:design} describes the design and introduces the architectural concepts of the \gls{sol} framework, while Section \ref{sec:implementation} provides implementation details.
We present the results of the evaluation of \gls{sol} in Section \ref{sec:evaluation}, covering both, measurements from our Android implementation as well as simulation results. Finally, in Section \ref{sec:conclusion}, we critically discuss implementation and performance issues, point to future work, and draw a conclusion.
\section{Related Work}
\label{sec:relatedwork}
Existing work in the field of security in decentralized networks
focuses mainly on the communication, providing secure routing protocols solutions \cite{schmittner2014scalable, li2016secure}, 
improving the fairness of the users in the network \cite{buchegger2002nodes}, increasing the robustness of such networks by detecting corrupt nodes \cite{shakshuki2013eaack, nadeem2014intrusion}, etc. 
Our work aims at proposing a solution to bootstrap security services, while integrating a scheme for authentication and key and trust management in a decentralized fashion. Quite a number of studies focusing on trust establishment \cite{seedorf2014decentralised} and key exchange using mobile devices have also been extensively analyzed \cite{farb2013safeslinger, shen2014secure}. Furthermore, there are also some proposals using existing security hardware on smartphones \cite{Busold:2013smart} and how these can be exploited to create different security levels. However, most of them either assume the existence of servers or lack an evaluation in both simulation and real devices.
Our work differs from the aforementioned solutions in that we propose, inspired by the approaches based on \gls{wot} model \cite{Capkun.2006, Cagalj.2006}, 
a more bare-bones implementation of decentralized authentication to make it more practical. Yet, our scheme offers cross-application support and easy integration into existing apps.
In our proposed scheme, each entity creates its public/private key and---after handshaking---issues certificates for its neighbors.
In addition, we consider the use of secure elements to provide a secure mechanism of key management locally in the devices. Finally, we investigate the performance by means of simulation, and we test our implementation on real devices. 
\section{Scenario and System Model}
\label{sec:assumptions}
In this section, we introduce the emergency response communication scenario that serves as a running example throughout the paper. We further present the general \gls{sol} system model as well as the adversary model.
\subsection{\textbf{Emergency Response Communication Scenario}}
Recent disasters \cite{HongKong.2014, Nepal.2015, Irma.2017} severely affected the information and communication capabilites of the population by damaging infrastructure, making communication systems unavailable, or knocking out the power. 
As a result, millions of people in need for help were literally left `in the dark', without ready-to-use access to backup power and striped even of basic means of communication.
A number of technical solutions exist to enable civilian volunteers and rescuers to build self-organized distributed wireless networks, thus enabling the population to communicate without relying on a centralized infrastructure. 
These networks build on user participation and leverage \gls{manet} or \gls{dtn} technology to facilitate message routing/forwarding/spreading in the affected area.
A typical assumption in emergency scenarios is that only honest users participate in establishing and running the network, and exis-ting solutions often forgo security means \cite{Alvarez:2016}.
As a result, users with malicious intend may limit or affect the communication, thus causing serious threats on the credibility and reliability of the data.
Throughout the rest of this paper, we will use the aforementioned emergency response scenario as a running example. 
However, this does not limit the generality of the proposed \gls{sol} framework.
\subsection{\textbf{System Model}}
We consider users owning mobile devices capable of direct device-to-device communication. 
In the following, we use the term entity to refer to the logical entity formed by an authorized user and her device.
Each device is imprinted with an identity, which is unique and unchangeable. 
For a smartphone this could be the \gls{imei}, a unique device fingerprint, etc.
Each device has a mechanism to discover devices in its proximity such as a one-hop neighbor discovery mechanism.
Applications running on the devices are assumed to be independent from each other, i.e., each application can define its own set of security requirements.
No prior trust relationships or security association between entities exist, i.e., no information or knowledge about other entities is stored on a device beforehand.
Centralized infrastructure to bootstrap security is unavailable.
By means of direct contact and with the users in the loop, pairwise trust relationship between entities can be established.
Note that we do not assume any technology for the data routing/forwarding/spreading, since \gls{sol} operates agnostic to such mechanisms.
\subsection{\textbf{Adversary Model}}
We assume adversaries that can act passively or actively as insiders, i.e., our adversary is a regular user of the network.
Adversary capabilities follow the Dolev-Yao assumptions \cite{dolev1983security}: the adversary is, hence, capable of active interception or mo-dification of traffic, she can fabricate and destroy messages, but is not able to break cryptographic primitives. 
The key goal of \gls{sol} being to bootstrap security, i.e., to provide authentication, key management and trust management services to the users, we define the main attack goals to be to disrupt these services.
In particular, this entails to impersonate other entities within the network.
\section{\textbf{S}ea \textbf{o}f \textbf{L}ights Framework Design and Architecture}
\label{sec:design}
In this section, we highlight the design concept and architecture of the proposed \gls{sol} framework.
We explain the technical realization in detail in Section \ref{sec:implementation}.
\gls{sol} is a framework that provides cross-application security services for
device-to-device communication settings. 
Our architecture comprises a key management as well as a trust management component, which are 
managed and developed as two independent elements. 
\subsection{\textbf{Decentralized Authentication}}
\label{subsec:protocols}
The proposed decentralized authentication is based on a simplified version of the \gls{wot} paradigm, 
where each mobile device generates its own public/private key pair and signs the public key of 
others devices. Similar to other \gls{wot} solutions, trust in \gls{sol} is determined by a trust level and a maximum certification path, the so-called \textit{\textbf{degree}}. We define the trust levels as follows. 
\begin{itemize}
\item \textbf{Ultimate (U)} for the owner, 
\item \textbf{Trusted (T)} for direct trust relations, where an object 
signed directly by \textbf{U} is trusted,
\item \textbf{Known (K)} for transitive relations (second degree and further), where
a object signed directly by \textbf{(T)} is known, or an object signed by \textbf{\textit{n}}-\textbf{K} 
is defined as known. \textbf{n} represents the minimum number of known signatures required to validate an unknown signature. Additionally, the degree defines if a transitive relation
can still be considered known. We do not set a fixed value of \textit{\textbf{n}} and \textit{\textbf{degree}}, but allow for user configuration. This enables to tune the scalability of the system to different use cases.
\end{itemize}
In our approach, trust management is carried out in two main steps, each implemented using a dedicated protocol. The \textit{handshake protocol} covers the bootstrapping and establishment of mutual trust, and the \textit{synchronization protocol} manages 
the unidirectional synchronization of the local trust repository. 

We denote a public key as $pk$, e.g., $pk[a]$ is the pu-blic key of \textbf{Device A}. Note that a certificate is represented as $signature[issuer, subject]$, where \textit{subject} is the device whose public key was signed, and \textit{issuer} is the device who signed the public key of the \textit{subject}, e.g., $signature[a,b]$ represents the certificate of $pk[b]$ issued by \textbf{Device A}.
\subsubsection{Handshake Protocol}
\begin{figure}[!ht]
	\centering
	\vspace{-0.15cm}
		\includegraphics[width=0.4 \textwidth]{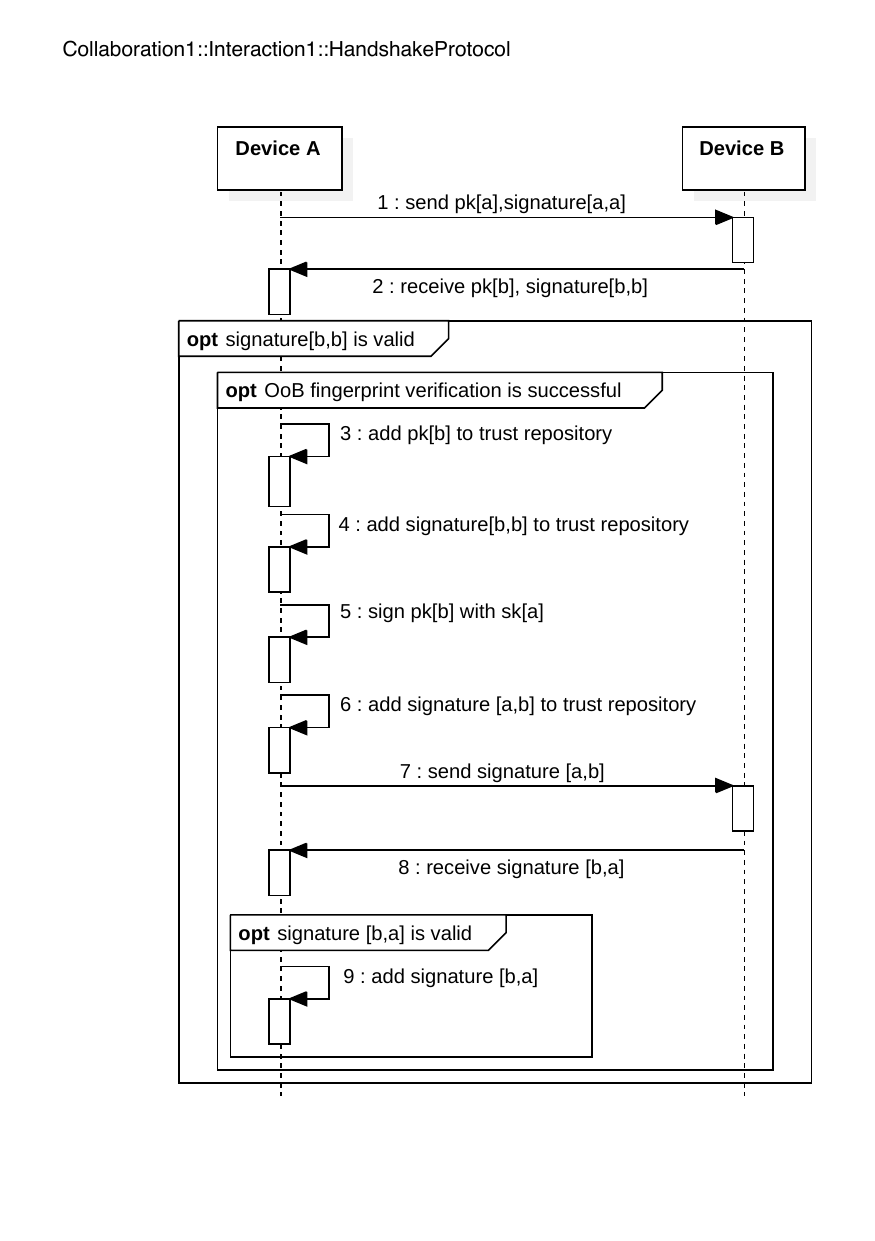}
		\vspace{-0.15cm}
	\caption{Bootstrapping trust process from \textbf{Device A}'s perspective}
	\label{fig:handshake}
	\vspace{-0.15cm}
\end{figure}

Figure \ref{fig:handshake} illustrates the data flow between two devices performing the handshake protocol.
It consists in the exchange of public keys and signatures between devices in proximity, 
in order to establish a direct trust relationship between two devices. 
The key verification is performed using existing Out-of-Band (OoB) verification methods (see \ref{sub:arc_trustlayer}). 
Since comparing all bytes of public keys can be tedious and susceptible to errors, we use a short representation of these called \textit{fingerprint}. 
In our approach, a fingerprint is the cryptographic hash value of any given public key. 
Once the OoB fingerprint verification is successful, the devices generate a certificate and assign a trust level according 
to the process mentioned previously. These certificates are then exchanged between devices and stored locally in their
repositories. 
\subsubsection{Synchronization Protocol} 
\begin{figure}[!ht]
	\centering
		\includegraphics[width=0.425 \textwidth]{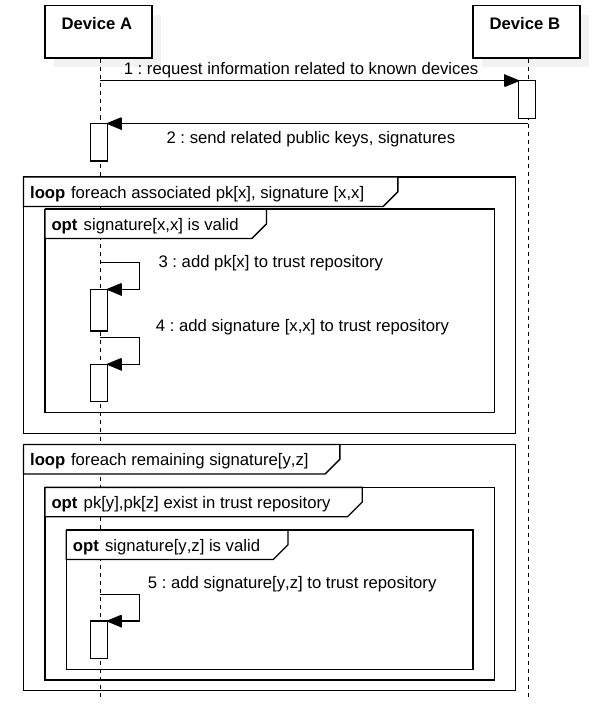}
	\vspace{-0.15cm}
	\caption{Unidirectional synchronization process from \textbf{Device A}'s perspective}
	\label{fig:synchronization}
		\vspace{-0.35cm}
\end{figure}

Once a trust relationship has been bootstrapped, the devices can obtain information about the 
transitive trust relations. Figure \ref{fig:synchronization} clarifies this process.
The synchronization of the trust repository between two devices operates as follows: \textbf{Device A} requests
from \textbf{Device B} information related to a set of known devices, \textbf{Device B} responds with the public keys 
and signatures related to the queried devices. Finally, \textbf{Device A} merges the information into 
its local trust repository.
\subsection{\textbf{Key Management}}
The key management component is responsible for the management and protection of the private authentication keys 
from misuse and key extraction.
The proposed solution is designed and implemented as a flexible solution, where the methods for the key 
management can be hardware- or software-based solutions. It depends on the methods supported
by the devices, e.g., keys can be stored in external \gls{nfc} tokens, TEE-based storage as AndroidKeyStore, etc. 
Note that irrespective of the method, the selected storage method requires a PIN, password 
or an additional unlock mechanism.
Our approach involves two group of keys: \textbf{(1)} the initial authentication key, which only aims to 
achieve authentication and trust between the devices; and \textbf{(2)} sub-keys. 

\textit{Sub-keys} are public/private key pairs, which can be used to provide additional security properties, 
e.g., confidentiality. These keys are generated by a third party app and authenticated with 
our framework using the initial authentication key. 
\subsection{\textbf{Architecture}}
\label{sub:architecture}
\begin{figure}[!ht]
	\centering	
	\vspace{-0.35cm}
		\includegraphics[width=0.5 \textwidth]{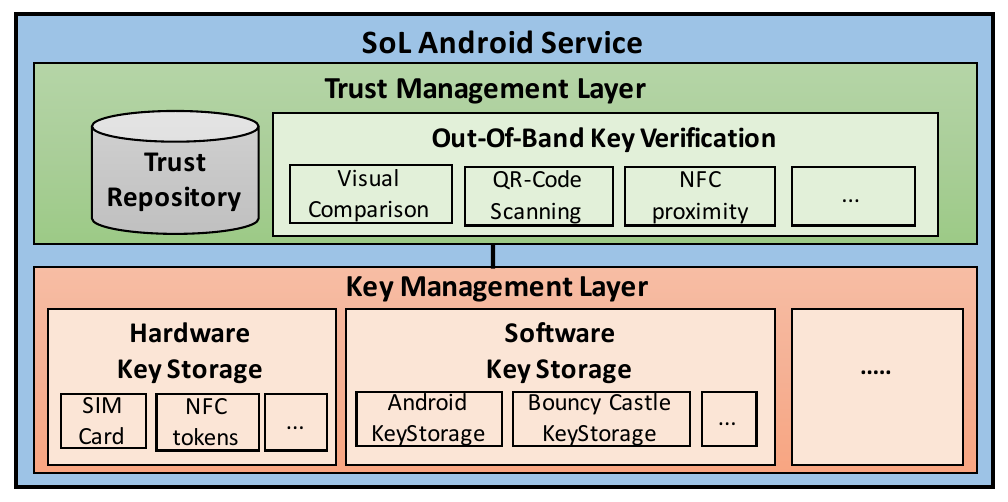}
			\vspace{-0.25cm}
	\caption{The \gls{sol} architecture is designed as an Android service}
	\label{fig:architecture}
	\vspace{-0.15cm}
\end{figure}

As shown in Fig. \ref{fig:architecture}, \gls{sol} is designed as a two-layer framework, which handles both trust and key management. It resides on the application layer of the Android 
architecture.
\subsubsection{Trust Management Layer}
The main task of this layer is the maintenance of the trust relationships on a device. 
It includes the management of the trust repository, controlling the data, e.g., existing public keys, certificates, sub-key certificates, as well as checking the validity and trustworthiness of incoming data.
Furthermore, this layer performs the protocols mentioned in \ref{subsec:protocols}: handshake and synchronization. 
\subsubsection{Key Management Layer}
The main goal of this layer is the initialization of the authentication private key, and the support of additional
operations where the initial authentication key is involved. It includes to sign keys, to issue certificates, etc. Moreover, this layer controls whether a storage is unlocked before performing any cryptographic operation. 
Whenever the storage is locked, a user interaction is needed for unlocking it, e.g., by introducing a password, lock pattern, PIN, etc. This layer also offers an interface to the upper Trust Management Layer.
\section{Implementation}
\label{sec:implementation}
We developed the \gls{sol} framework for the Android platform and implemented it as a bound service\footnote{https://developer.android.com/guide/components/bound-services.html} 
(=\textit{\gls{sol} service}) 
by using Android Interface Definition Language\footnote{https://developer.android.com/guide/components/aidl.html}.
The \textit{\gls{sol} service} is encapsulated inside a standalone Android application running in its own process. It implements the 
trust and key management layers defined in \ref{sub:architecture}.
In the following, we discuss the most important implementation details for all \gls{sol} components.
\subsection{\textbf{Trust Management Layer}}
\label{sub:arc_trustlayer}
We use \textbf{fingerprints} and \textbf{key IDs} to identify a longer key with a short representation. 
The fingerprint is calculated using SHA-256, and the key ID is derived from the 64 LSBs of the public key. 
We create a directory that contains all data concerning trust relations for each known device (\textit{=subjects}). 
The directory's name is the hexadecimal representation of the fingerprint for a subject public key. 
The directory contains the subject public key, signatures over the subject public key, as well as all sub-keys and 
their respective certificates attached to the subject.
All these data are serialized and stored persistently as Base64-encoded files. The generated files are located in
the application private directory.

\begin{figure}[!ht]
	\centering
	\vspace{-0.15cm}
		\includegraphics[width=0.425 \textwidth]{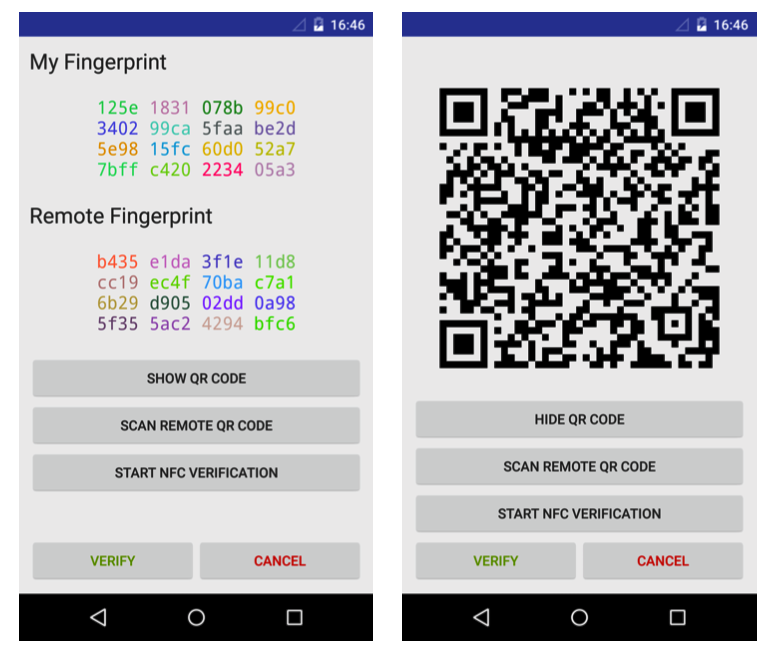}
		\vspace{-0.20cm}
	\caption{Example of two key verification methods implemented in \gls{sol}}
	\label{fig:verification}
	\vspace{-0.15cm}
\end{figure}

This layer is also responsible for the key verification. After successful completion of the handshake protocol, the key exchanged 
in this protocol needs to be authenticated. 
Our framework allows for easy extensibility by facilitating the new implementation, extension, or replacement of key authentication modules.
Currently, we have implemented the following existing authentication mechanisms:
\begin{itemize}
\item \textbf{Visual comparison: } The remote and the own fingerprint are color-coded and displayed to the users as shown in Fig. \ref{fig:verification}.
\item \textbf{Scanning a QR code: } Both fingerprints are encoded in Base64 and 
encapsulated in a QR code as shown in Fig. \ref{fig:verification}. The user is required to scan the remote QR code. The scanning is performed using the 
ZXing (Zebra Crossing) project \cite{ZXing}.
\item \textbf{Using \gls{nfc} technology: }  When the devices are in close proximity, they exchange the fingerprints 
automatically. 
\end{itemize}
\subsection{\textbf{Key Management Layer}}
The Key Management Layer implements the necessary mechanisms to provide a modular and
flexible key storage solution. Thus, this layer is split into the following sub-modules:
\begin{itemize}
\item \textbf{SoftwareKeyManager: } This sub-module supplies a software-based keystore solution 
developed for Android versions before 4.3. The keystore is based on the Bouncy Castle library. 
The files generated in this module are protected using a user-provided PIN or pass-phrase. 
\item \textbf{AndroidKeyManager: } The AndroidKeyManager is the solution for Android versions from 4.3 on. This module utilizes the official Android API keystore, which introduces an application private credential storage concept, but also 
(if available) enables additional security by offering support for hardware-based solutions. 
\item \textbf{HardwareKeyManager: } The HardwareKeyManager represents an additional abstraction layer 
for the key storage. It is the basis module for all hardware-based solutions, as all these solutions employ a similar
protocol based on certain \gls{apdu} commands to communicate with Java Card Applets.
Our current implementation covers three hardware-based solutions, which include the key storage and also perform the required signature operations: 
\textbf{(1)} \textit{SmartcardManager} handles the communication with \gls{nfc} smart cards.
\textbf{(2)} \textit{SeekManager} supports the connection to available rea-ders, e.g., UICC. The Seek Manager manages the communication with the existing SEEK for Android framework \cite{Seek}.
\textbf{(3)} \textit{YubiKeyManager} permits the communication with YubiKey NEO hardware tokens \cite{YubiKey}. 
\end{itemize} 
\subsection{\textbf{Solution-dependent Settings}}
In our implementation, we abstract the network layer tasks, that is, neighbor discovery and data transmission. This abstraction allows to replace the ad-hoc communication with another technology at any time.
In our proof-of-concept, we used Wi-Fi Direct as the ad-hoc communication technology.\\
Moreover, we prioritize the selection of the key manager according 
to the solutions supported by the device. 
\subsubsection{Configurable Properties}
We define several properties as configurable:
\begin{itemize}
\item \textbf{maxdegree:} defines the maximum number of transitive relations that can still be considered valid. 
\item \textbf{numknown:} fixes the number of required known signatures to validate an
unknown signature. 
\item \textbf{maxsubkeys:} determines the maximum number of sub-keys that an application can register.
\item \textbf{signaturealgorithm:} represents the selected signature algorithm, e.g., RSA or ECDSA.
\end{itemize}
\subsubsection{Choosing the most suitable key manager}
Our selection prioritizes the hardware-based solutions. During the initia-lization of the \gls{sol} service, 
it checks whether any reader is available. If it exists, the \textit{SeekManager} is chosen. If it does not exist, 
we ask the user if she wants to utilize other supported hardware-based method, e.g., \textit{SmartcardKeyManager}
or \textit{YubiKeyManger}. Otherwise, we examine the running Android version and automatically select the suitable 
software-based module.
\subsection{\textbf{Integration \gls{sol} Android service by third party applications}}
We provide a proxy library (=\textit{\gls{sol} library}) that allows an app to 
communicate with the \gls{sol} service in a simple and direct fashion. 
\begin{figure}[!ht]
	\centering
	\vspace{-0.20cm}
		\includegraphics[width=0.475 \textwidth]{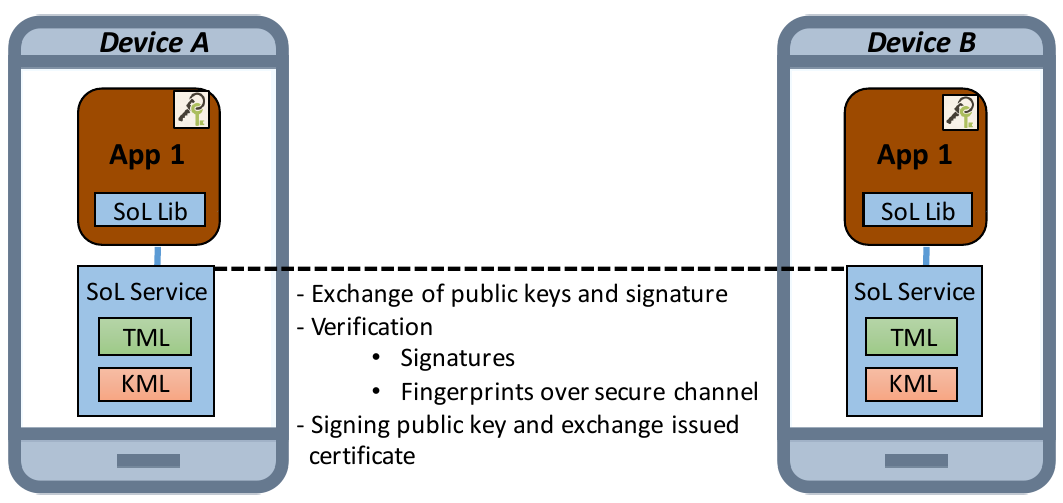}
		\vspace{-0.20cm}
	\caption{\gls{sol} Android service performing the handshake protocol}
	\label{fig:dataflowhsp}
	\vspace{-0.35cm}
\end{figure}

\begin{figure}[!ht]
	\centering
\vspace{-0.30cm}
		\includegraphics[width=0.475 \textwidth]{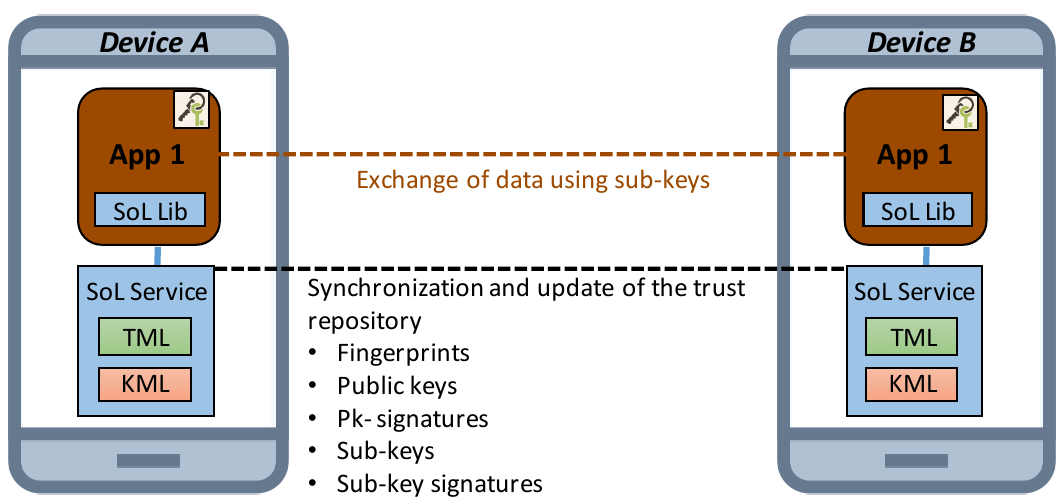}
		\vspace{-0.20cm}
	\caption{\gls{sol} Android service performing the synchronization protocol}
	\label{fig:dataflowsp}
	\vspace{-0.15cm}
\end{figure}
The main tasks of this proxy library include:
\begin{itemize}
\item Validate the installation, availability and successful initialization of the \gls{sol} service. 
\item Trigger the service to start the handshake protocol.
\item Check existing trust relationships with a neighbor. 
\item Retrieve additional information about a neighbor.
\item Request and register app specific sub-key certificates.
\item Return the sub-keys associated to a specific fingerprint.
\end{itemize}
To clarify the use of our \gls{sol} framework, we assume the following scenario. 
Let \textbf{App 1} be a chat app that wants to provide a secure communication between devices in a
decentralized manner. For doing so, it creates a sub-key using an asymmetric algorithm scheme. 
To benefit from our framework, \textbf{App 1} uses the \textit{\gls{sol} library} 
to register the generated public key in our framework. 
Hence, the \gls{sol} service signs the key and
issues a certificate. Finally, the certificate and the public key are stored in our local trust repository.
Then, the \gls{sol} service takes care of the distribution and synchronization of the public key as well as its certificate.
It implies that multiple apps running on a smartphone neither need to build nor to maintain their own trust repository. Figures \ref{fig:dataflowhsp} and  \ref{fig:dataflowsp} show the data flow between devices in the aforementioned scenario. 
\section{Evaluation}
\begin{table}[!ht]
\renewcommand{\arraystretch}{0.49}
	\caption{Experiment Settings}
	\label{tab:onesettings}
	\centering

	\resizebox{0.49\textwidth}{!}{
		\begin{tabular}{lll}
		\hline
		\hline
	
		\hline
		\\
		\textbf{\textit{Scenario}}  &
		Dimensions \textit{w x h} & 
		3000 x 3000 [m] \\   
		\\
		
		 &
		Simulation duration & 
		12 [h], i.e., 720 [min] or 43200[s] \\   
		
		\\
		
		 &
		Number of nodes & 
		120 \\   
		
				\\		
		 &
		Experiment & 
		6 ( 3 trust degree x 2 signature algorithm) \\   
		
		\\		
		 &
		Runs & 
		5 per experiment \\   
		
		\\
		\hline
		\\
		
		\textbf{\textit{Mobility}}  &
		Model & 
		RWP \\   
	
		\\		
		 &
		Speed & 
		0.5, 1.5 [m/s] \\   
	
		\\
		\hline
		\\
		
		\textbf{\textit{Routing}} &
		Algorithm & 
		DirectContact \\   
		
		\\
		
		 &
		Buffer size & 
		20 [MB] \\   
		
		\\
		\hline
		\\
		
		\textbf{\textit{Communication}} &
		Transmit speed & 
		2 [Mbps] \\   
		
		\\
		 &
		Transmission range & 
		10 [m] \\   
		
		\\
		\hline
		\\
		
		\textbf{\textit{Trust management}}  & 
		Maximum trust degree &
		1 (direct) - 3 \\ 
			
		\\
		\hline
		\\
		\textbf{\textit{Key management}}  &
    		Number of Sub-keys & 
    		3 \\		
		\\
		 & 
		 Size per Sub-key  & 
		 4096 bit \\ 
		 \\
		
		&
		Signature algorithm & 
		RSA (2048 bit) \\ 

		\\
          & 
          & 
          ECDSA (256 bit) \\ 
          \\
		\hline
		\hline
		
		\end{tabular}
		}
		\vspace{-0.38cm}
\end{table}
\label{sec:evaluation} 
The goal of the evaluation is twofold. First, we investigate the performance and scalability aspects of the trust management and the key management layer by means of simulation using the \gls{one}\cite{keranen2009one}. 
Second we demonstrate and test our implementation on real devices to show its feasibility and test the computational performance of the key management part (see also \cite{sinha2013performance} for existing performance studies of the employed algorithms).
\subsection{\textbf{Trust Management}}
Detailed simulation settings for the \gls{one} are provided in Table \ref{tab:onesettings}. 
We analyze four evaluation metrics for this layer: the propagation of trust relations, memory consumption, 
bandwidth consumption and computational overhead.
The maximum degree of transitive trust relations varies from [1,3]. Each node starts with a maximum of 3 sub-keys. Each experiment runs for 12 hours (720 minutes) and nodes exchange their data every 10 seconds, if in proximity. 
The plots show averages over 6 different experiments: one per transitivity degree (degrees 1 to 3) using 2 possible signature algorithms (RSA or ECDSA). Each run is seeded with numbers from the interval [1,5], resulting in a total 30 runs. We use BouncyCastle JCA for signing and key generation operations. We show average values and omit the confidence intervals, which are sufficiently small and would hamper readability.
\subsubsection{Trust relationships}
\begin{figure*}[b]
   \centering
   \vspace{-0.3cm}
   \subfloat[Direct relation - First degree]{
       		\includegraphics[width=0.315 \linewidth]{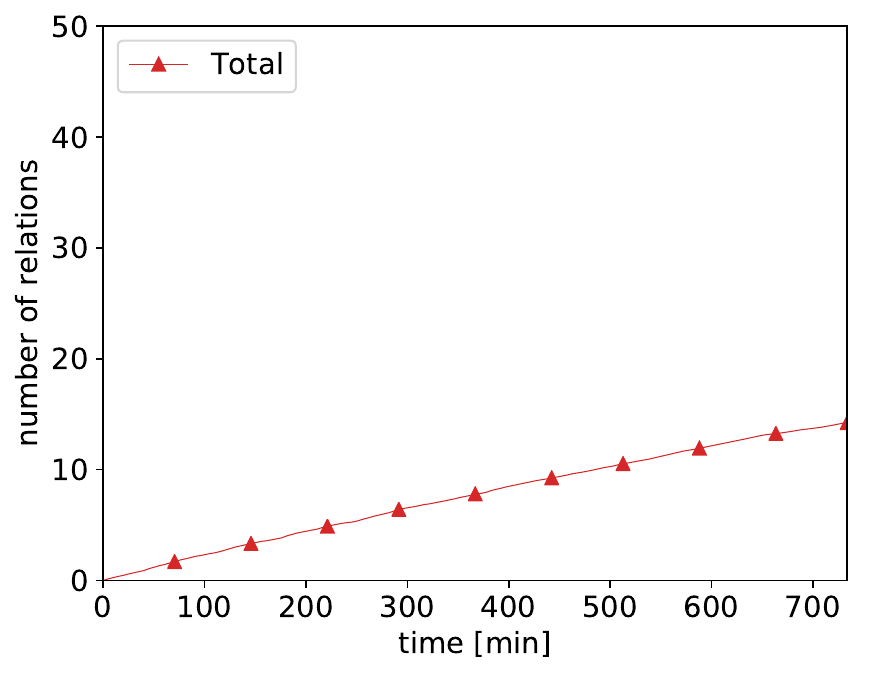}}
        \label{fig:d1}
    \subfloat[Indirect relations : Second degree]{
       		\includegraphics[width=0.315 \linewidth]{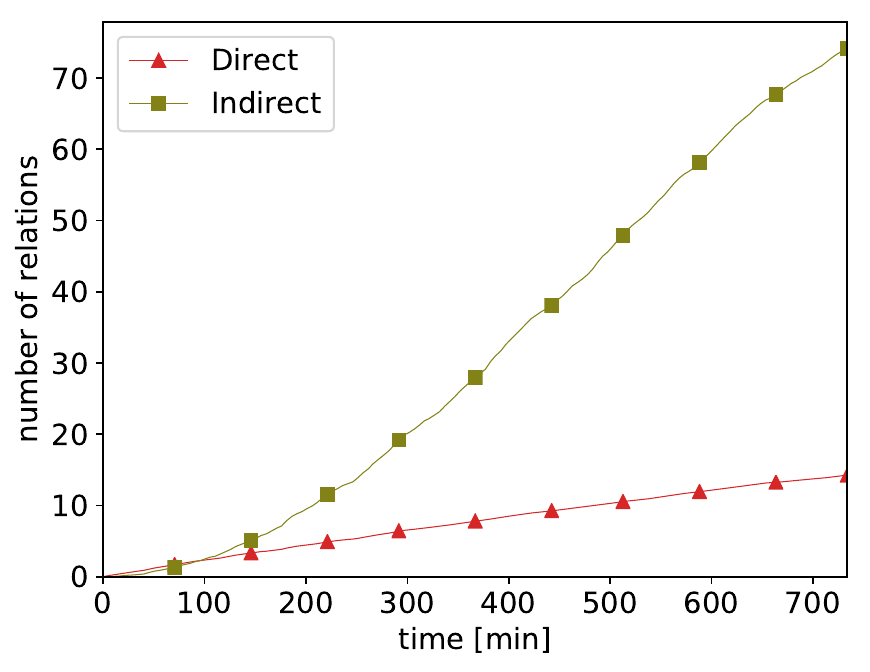}}
        \label{fig:d2}
   \subfloat[Indirect relations : Third degree]{
       		\includegraphics[width=0.315 \linewidth]{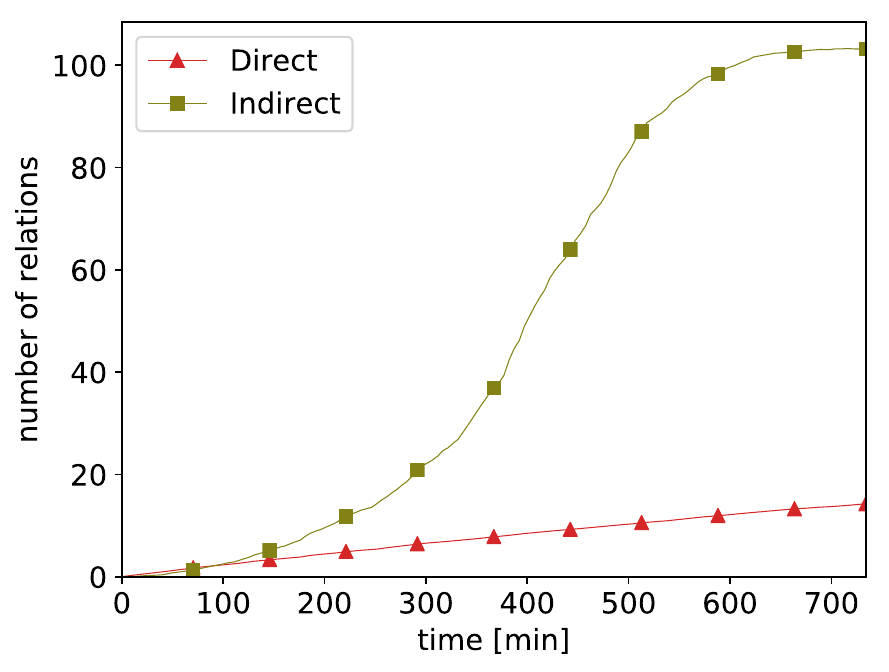}}
        \label{fig:d3}
    \caption{Propagation of direct and indirect trust in the network}\label{fig:trust_degree}
      \vspace{-0.2cm}
\end{figure*}

Figure \ref{fig:trust_degree} shows the number of direct trust relationships as well as the implicit relationships. 
While implicit relationships increase exponentially, the direct trust indicates a linear property. 
The number of direct trust relationships remains almost the same and it does not depend on the maximum
certification path. But rather, it varies according to the number of performed handshakes between devices. 

\subsubsection{Memory and bandwidth consumption}
For this expe-riment, we determine the file size required for public keys, signatures, sub-keys and sub-keys signatures in the repository.
\begin{table}[!ht]
\renewcommand{\arraystretch}{0.49}
	\caption{Proof-Of-Concept Settings}
	\label{tab:proofofconcept}
	\centering

	\resizebox{0.49\textwidth}{!}{
		\begin{tabular}{lll}
		\hline
		\hline

		\hline
		\\
		\textbf{\textit{Scenario}}  &
		Signature algorithm & 
		RSA (2048 bit) \\ 

		\\
          
          &
          & 
          ECDSA (256 bit) \\ 
		 \\
		 
		 &
		Devices & 
		1 ThinkPad X220, 8GB RAM, Ubuntu 64 bit \\   
		
		\\
		
		&
		 & 
		1 Google Nexus 5, Android version 6.0\\   
		
		\\
		&
		 & 
		1 YubiKey NEO (NFC token) \\   
		
		\\
		\hline
		\\
		
		\textbf{\textit{Procedure}}  &
		I.   Key Generation & 
		2 key-pairs (\textit{kp1,kp2}) \\   
	
		\\		
		 &
		 & 
		1 invalid signature \\   
	
		\\
		 &
		II.  Issue signatures & 
		1000 by \textit{kp1} \\   
		
		\\
		 &
		 & 
		200 by \textit{kp2} \\ 
	
		\\
		 & 
		 III. Verifications &  
		 1000 (valid) \\
		 
		 \\
		 & 
		 &  
		 200 (invalid) \\

          \\
		\hline
		\hline
		
		\end{tabular}
		}
				\vspace{-0.38cm}
\end{table}

\begin{figure*}[!bth]
    \centering
    \subfloat[ECDSA - First degree]{
       		\includegraphics[width=0.315 \linewidth]{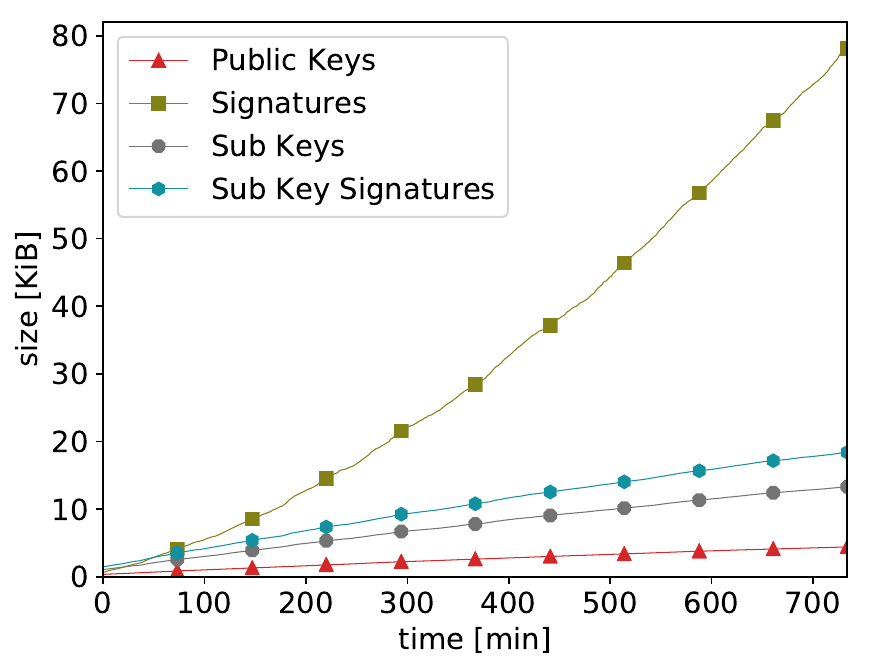}}
        \label{fig:eccmemd1}
   \subfloat[ECDSA - Second degree]{
       		\includegraphics[width=0.315 \linewidth]{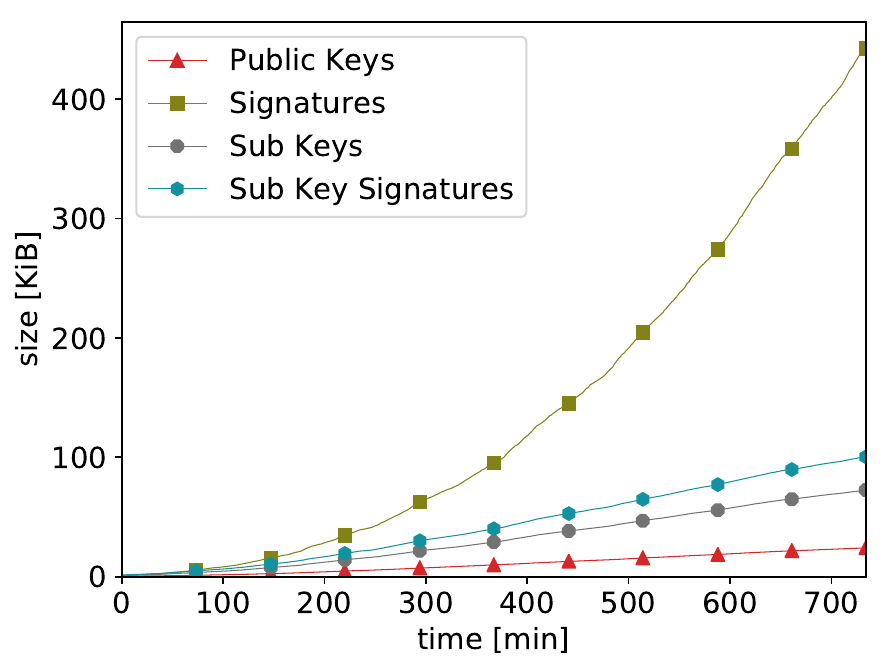}}
        \label{fig:eccmemd2}
          \subfloat[ECDSA - Third degree]{
       		\includegraphics[width=0.315 \linewidth]{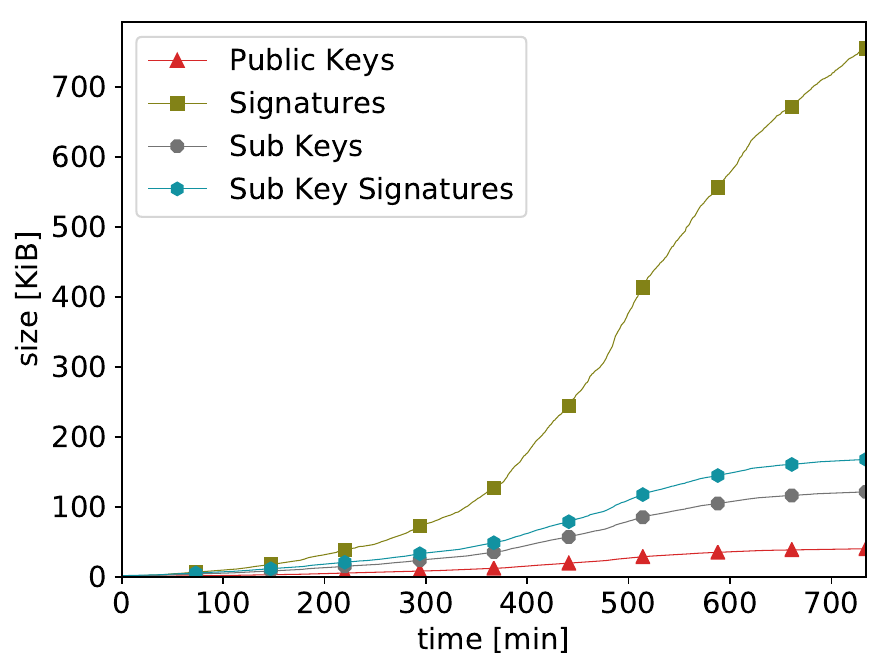}}
        \label{fig:eccmemd3}
        \subfloat[RSA - First degree]{
       		\includegraphics[width=0.315 \linewidth]{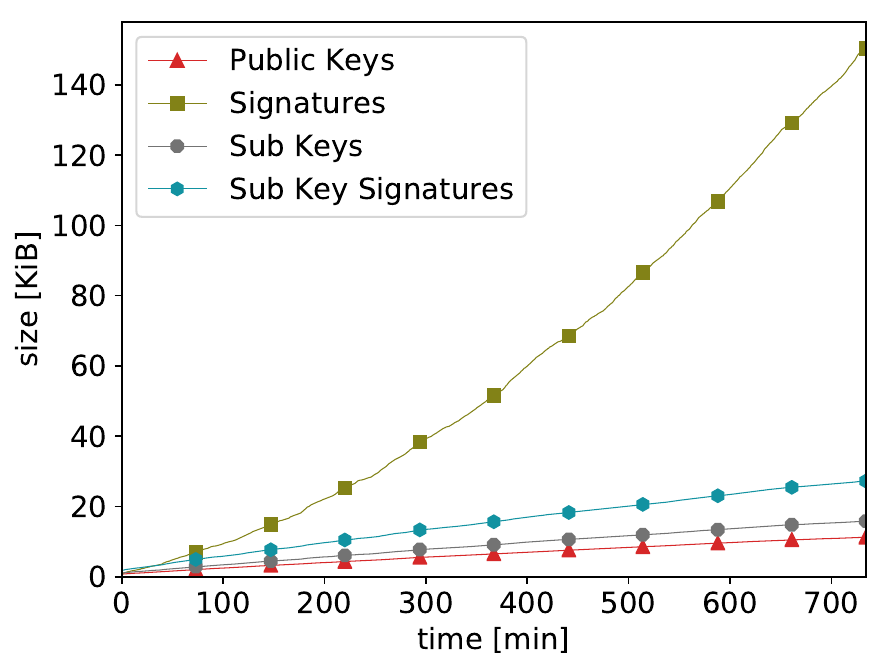}}
        \label{fig:rsamemd1}
         \subfloat[RSA - Second degree]{
       		\includegraphics[width=0.315 \linewidth]{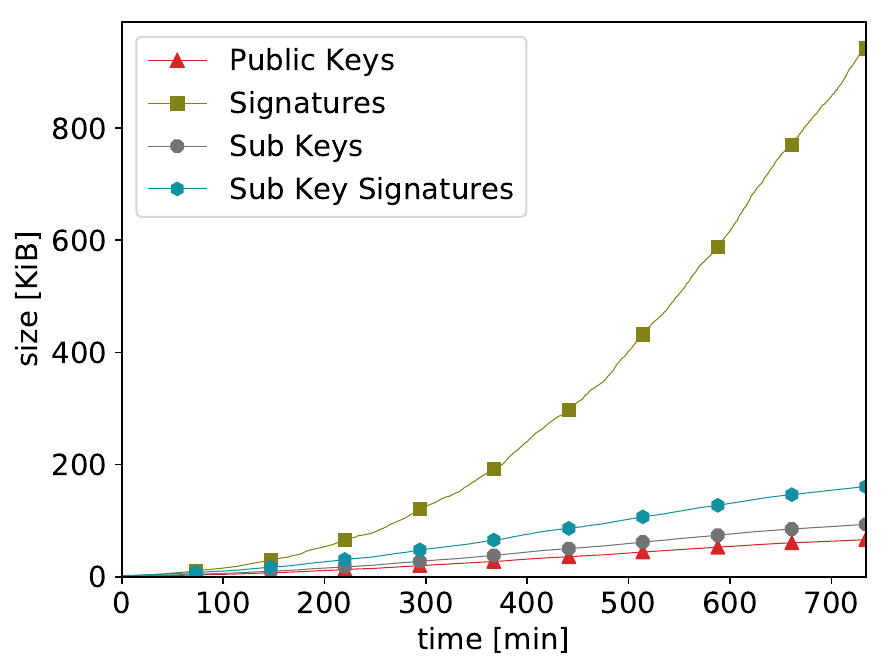}}
        \label{fig:rsamemd2}
              \subfloat[RSA - Third degree]{
       		\includegraphics[width=0.315 \linewidth]{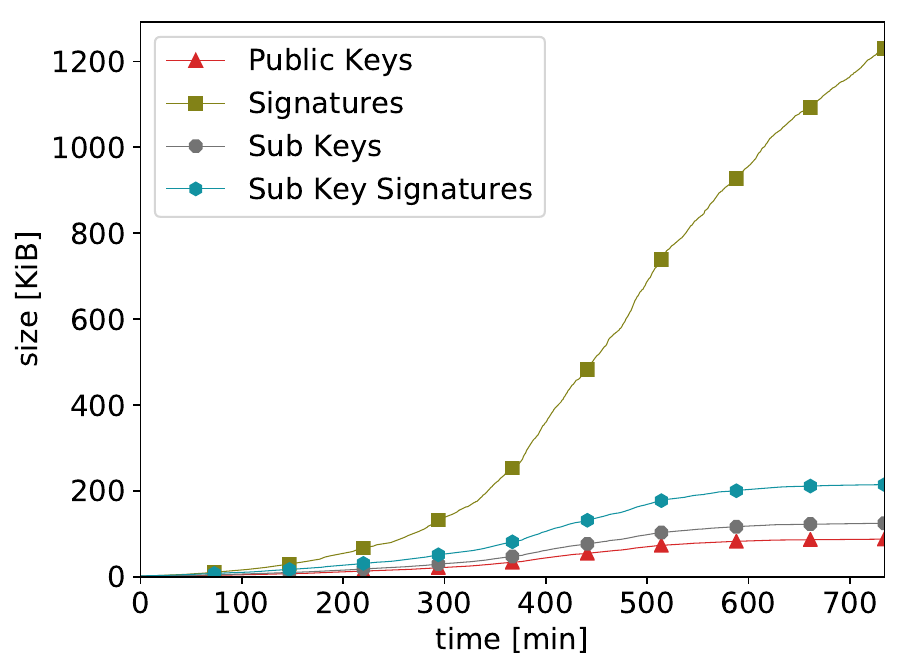}}
        \label{fig:rsamemd3}
    \caption{Comparison of memory consumption using different cryptographic algorithms }
    \label{fig:memory}
\end{figure*}

As depicted in Figure \ref{fig:memory}, the memory consumption is directly influenced by the selected trust 
degree as well as the signature algorithm. 

\begin{figure*}[b!ht]
    \centering
    \subfloat[Handshake phase ]{
       		\includegraphics[width=0.325 \linewidth]{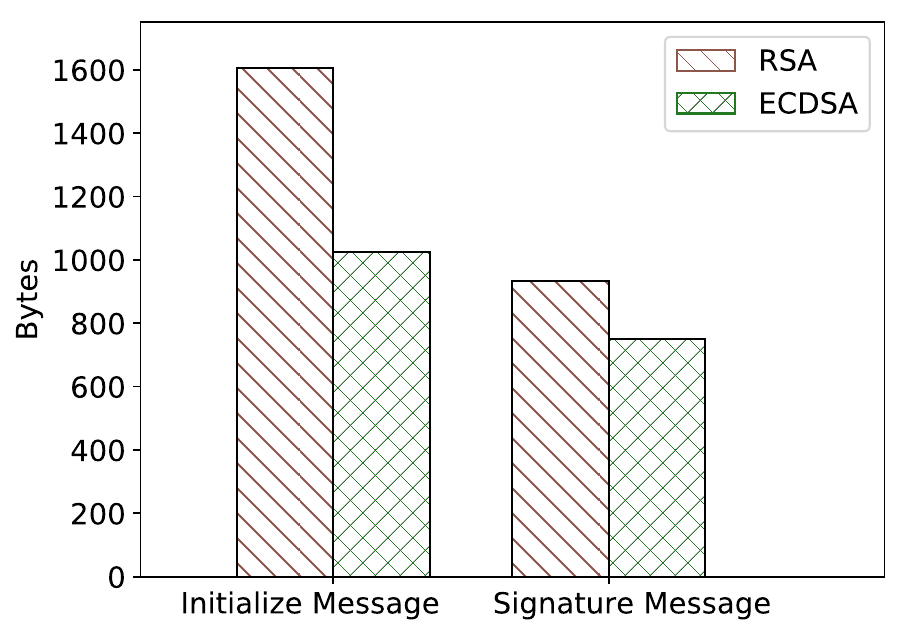}}
        \label{fig:hs_sync}        
    \subfloat[First degree]{
       		\includegraphics[width=0.3 \linewidth]{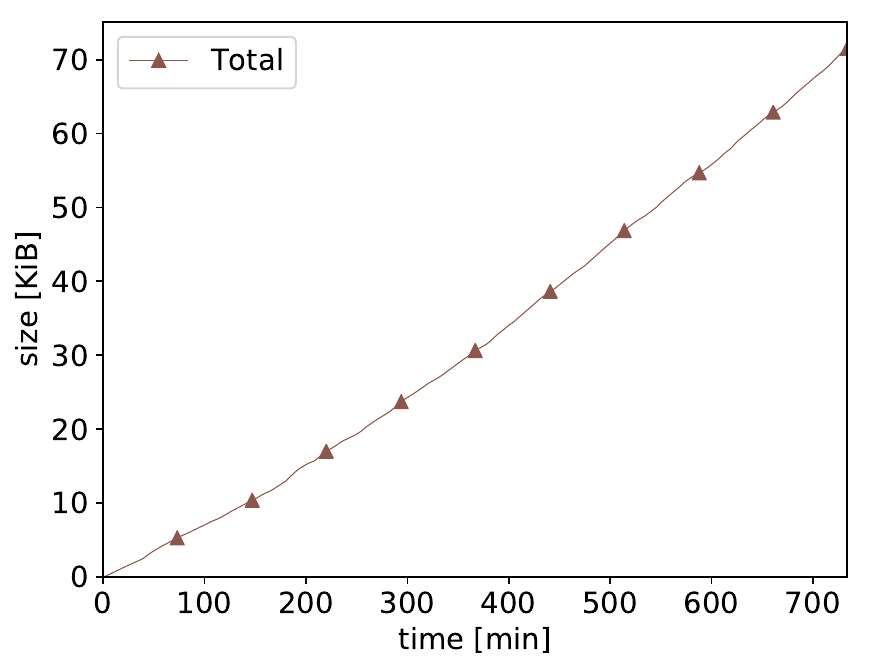}
       		\includegraphics[width=0.2 \linewidth]{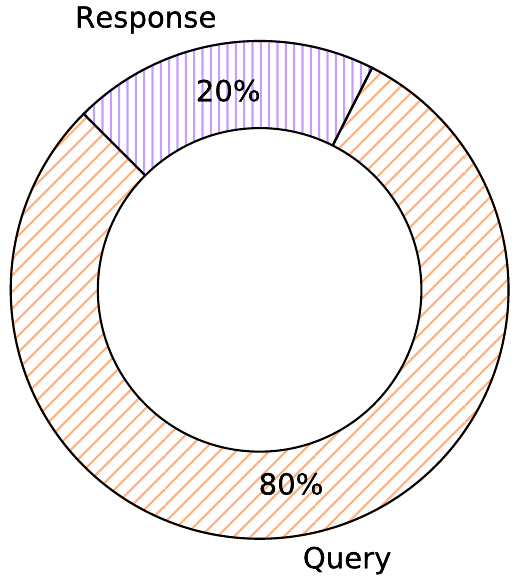}}
        \label{fig:bandwidth2}
    \caption{Bandwidth usage during: (a) handshake phase, (b) synchronization phase: total (left) - query and response operations (right)}
    \label{fig:bandwidth_phase}
\end{figure*}

\begin{figure*}[b!ht]
    \centering
     \subfloat[Second degree]{
       		\includegraphics[width=0.28 \linewidth]{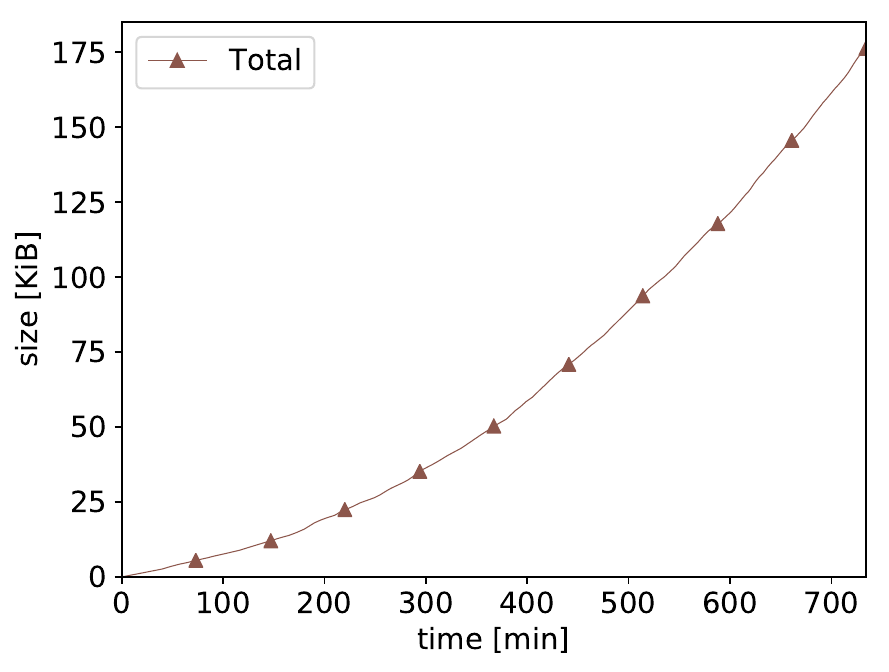}
       		\includegraphics[width=0.18 \linewidth]{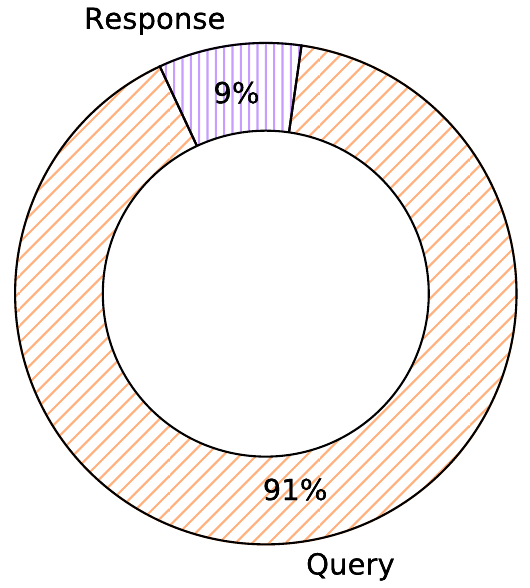}}
        \label{fig:bandwidth2}      
            \subfloat[Third degree]{
       		\includegraphics[width=0.28 \linewidth]{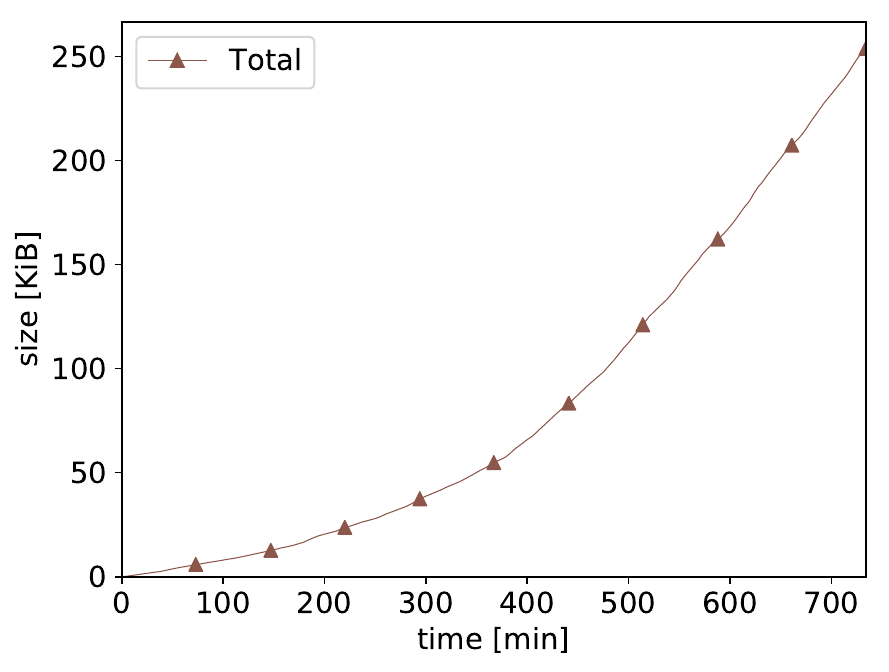}
       		\includegraphics[width=0.18 \linewidth]{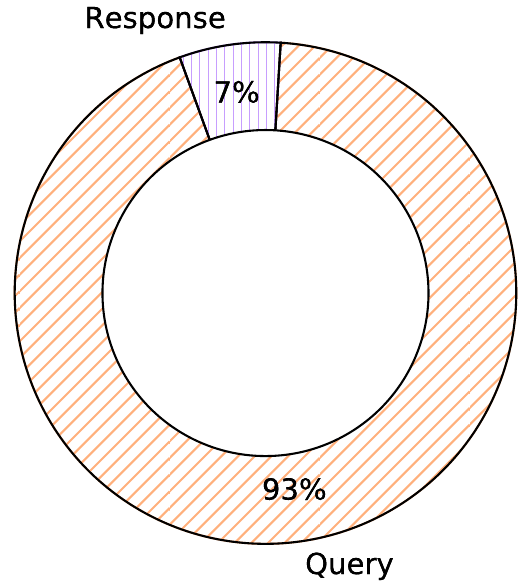}}
        \label{fig:bandwidthpercent3}
    \caption{Bandwidth usage during synchronization phase: total (time plots) and query and response operations (pie plots) }
    \label{fig:bandwidth_percent}
\end{figure*}

First, our results confirm the existing findings regarding both algorithms: RSA exhibits higher memory usage than ECDSA for the generation of the primitives public key, sub-keys and the signatures. 
As the number of collecting signatures of known devices increases with each neighbor encounter, 
the storage space in the repository is mainly occupied by signatures. 
This in turn implies a considerable memory overhead, thus each node collects and stores signatures according to the trust degree selected.\\
Figures \ref{fig:bandwidth_phase} and \ref{fig:bandwidth_percent} show the bandwidth overhead and usage required for the handshake and the synchronization phase. Although, the bandwidth usage is constant during the handshake protocol, it 
increases rapidly in the synchronization phase depending on the selected maximum degree of trust relations. \\
Furthermore, if we split the data transferred during the synchronization phase into \textbf{query} and 
\textbf{response} operations, we notice that query operations account for the overwhelming part of usage bandwidth during the synchronization, which further increases with increasing trust degree.
This is a very important result: on one hand, a trust degree higher than one is important to scale up the \gls{wot} faster, on the other hand such higher degree burdens the network and, thus, may impact the expansion of the \gls{wot} due to overload situations. In Section \ref{sec:conclusion} we suggest possible optimization mechanisms to minimize this issue. 
\subsubsection{Computational Overhead}
\begin{figure*}[!t]
    \centering
    \subfloat[Signature - all]{
       		\includegraphics[width=0.22 \linewidth]{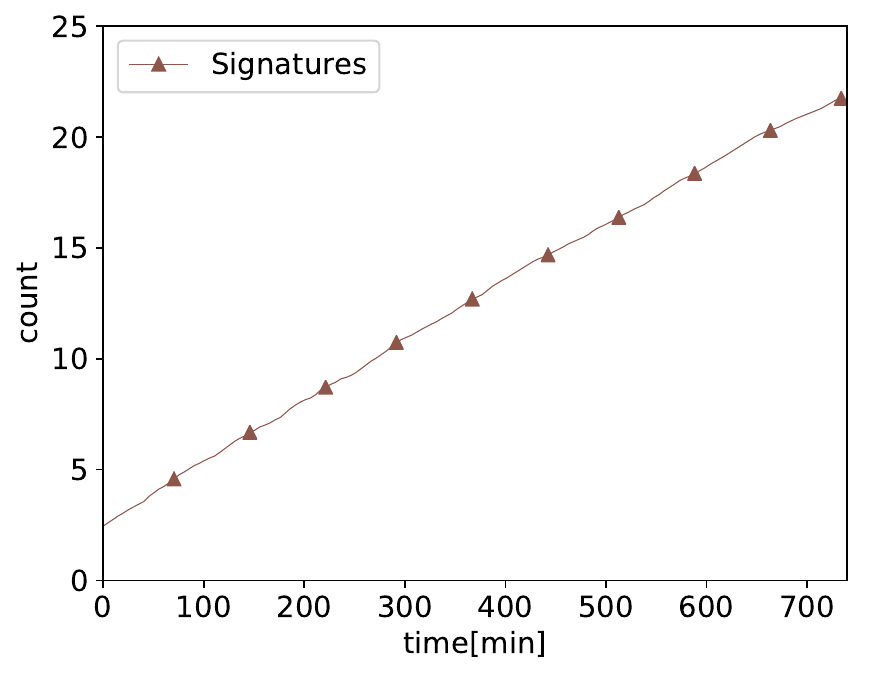}}
        \label{fig:sig_all}        
    \subfloat[Verification - First degree]{
       		\includegraphics[width=0.245 \linewidth]{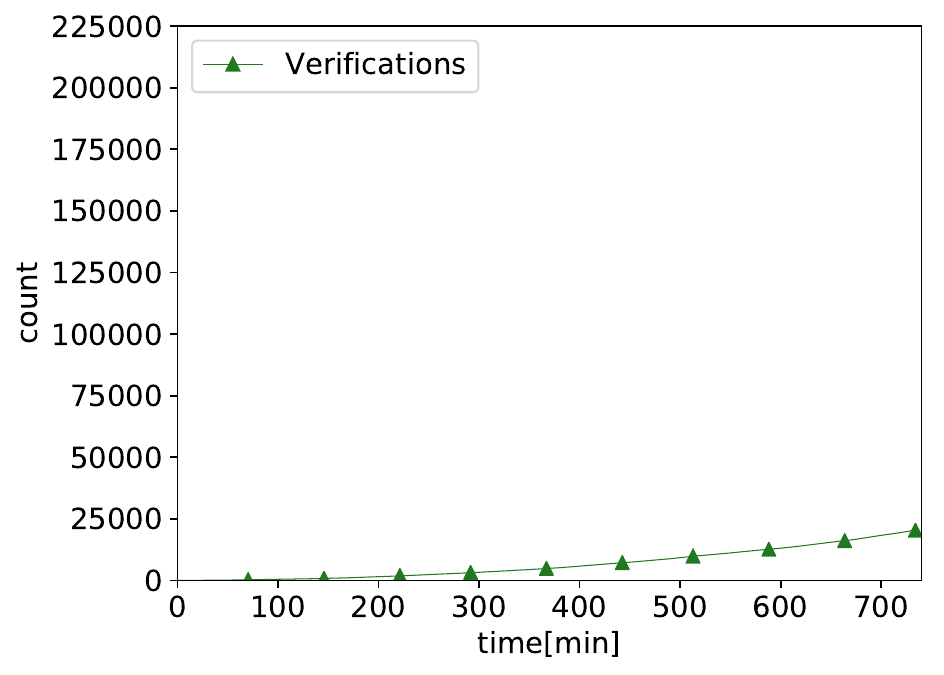}}
        \label{fig:verificationd1}      
   \subfloat[Verification - Second degree]{
       		\includegraphics[width=0.245 \linewidth]{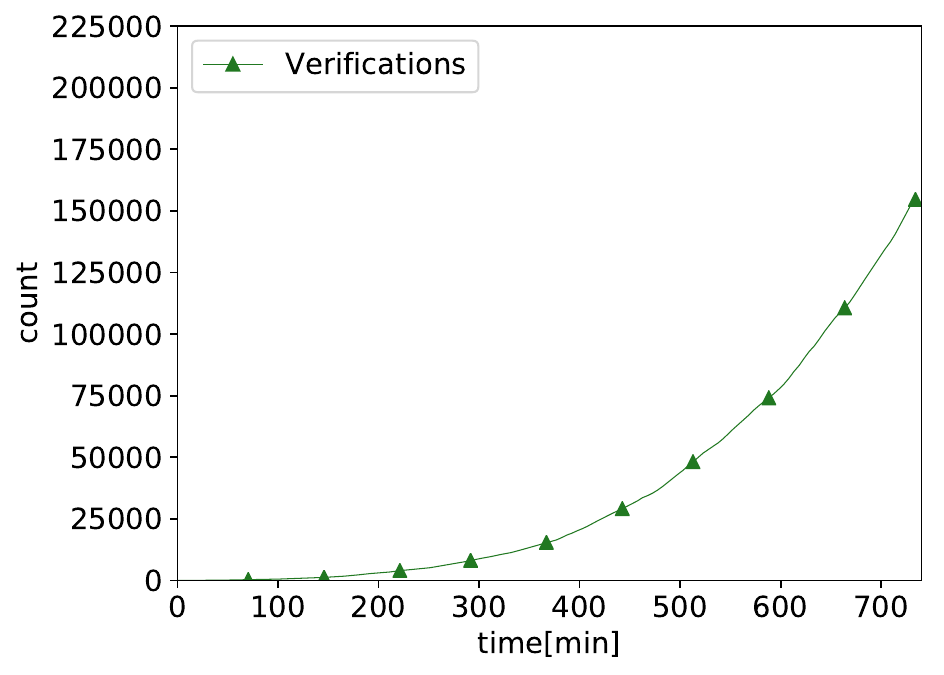}}
        \label{fig:verificationd2}
           \subfloat[Verification - Third degree]{
       		\includegraphics[width=0.245 \linewidth]{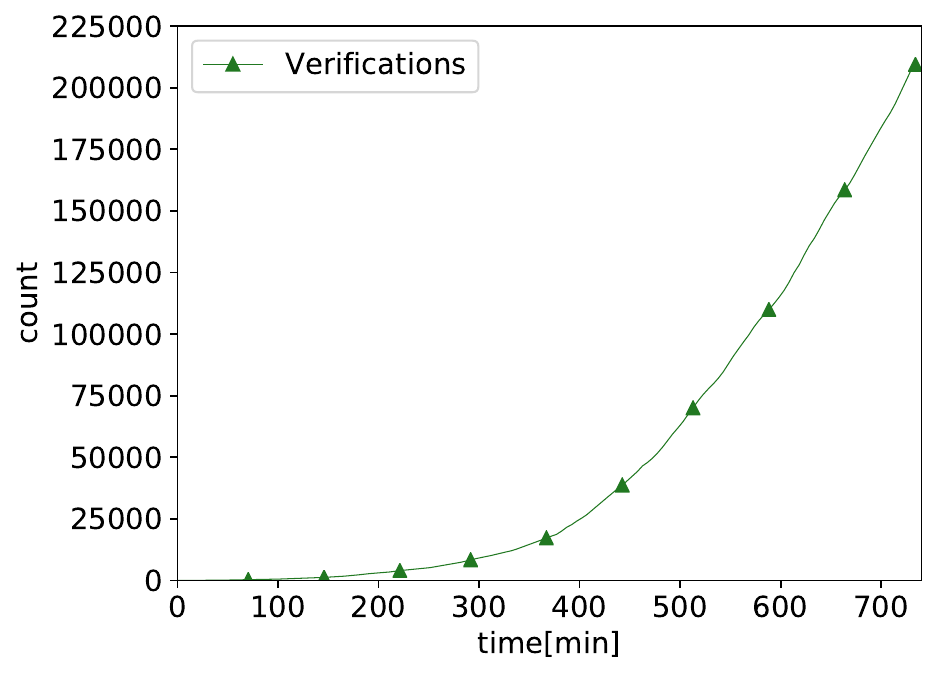}}
        \label{fig:verificationd3}
    \caption{Computational overhead: total numbers of signing and verification operations during the simulation}
    \label{fig:sig_ver}
\end{figure*}

We analyze the computational overhead of our solution based on numbers of operations realized during the simulation. 
Because these number of operations is the same for both signature algorithms, we do not separate the results into ECDSA and RSA. 
As shown in Figure \ref{fig:sig_ver}, verification represents the most significant operation performance wise.
Its growth is exponential and directly associated to the trust degree.
\subsection{\textbf{Key Management Layer}}
A proof-of-concept of \gls{sol} was implemented on Android-based smartphones to demonstrate the feasibility of our framework on real devices. 

The evaluation of this layer covers the two supported cryptographic algorithm, and compares both in 
terms of efficiency and performance. Additionally, we compare hardware- and 
software-based key storage mechanisms. \\
Detailed experiment settings are provided in Table \ref{tab:proofofconcept}. 
\begin{figure*}[b!ht]
    \centering
    \subfloat[Key generation]{
       		\includegraphics[width=0.325 \linewidth]{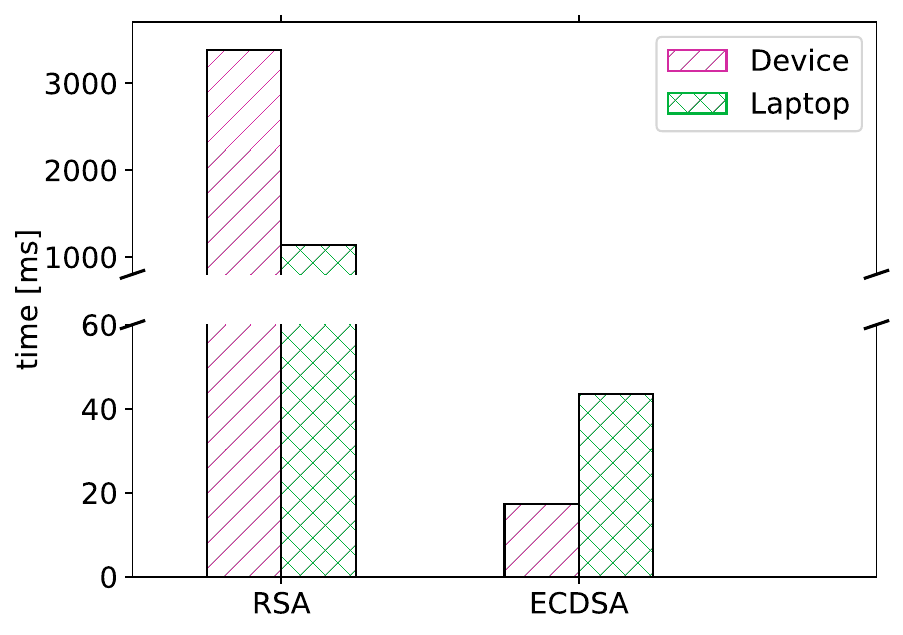}}
        \label{fig:generation}
   \subfloat[Signing]{
       		\includegraphics[width=0.3175 \linewidth]{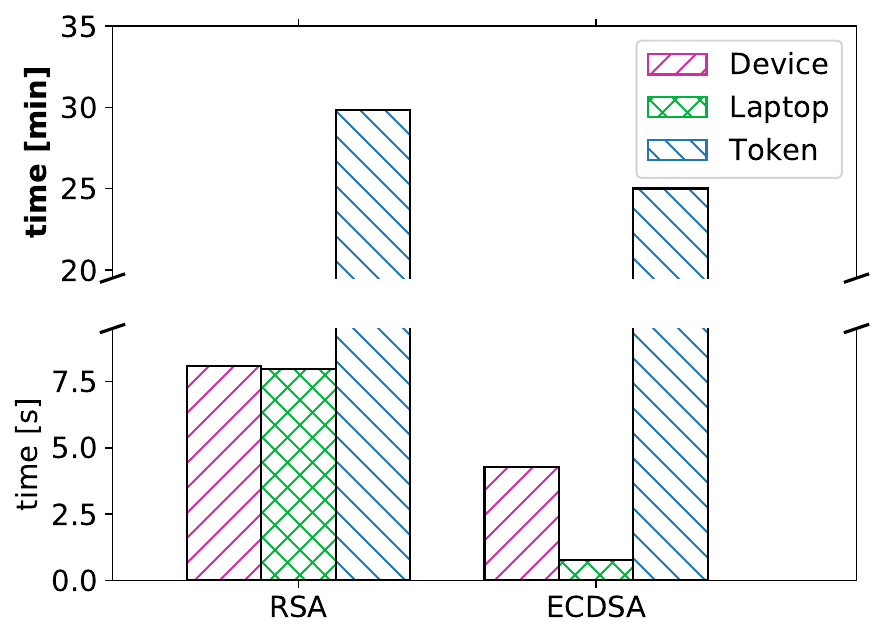}}
        \label{fig:signing}
           \subfloat[Verification]{
       		\includegraphics[width=0.325 \linewidth]{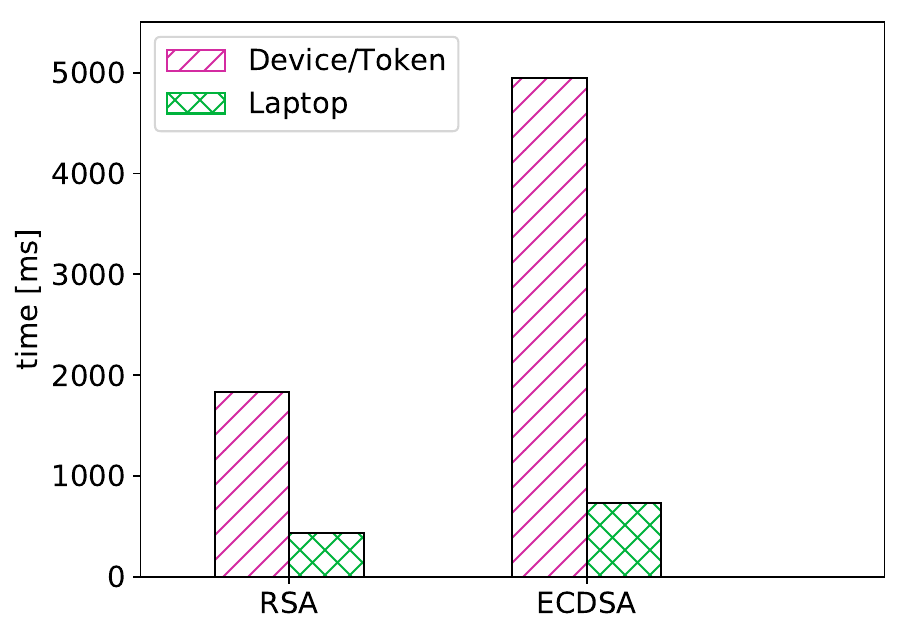}}
        \label{fig:verification_poc}
    \caption{Performance comparison between RSA and ECDSA executing cryptographic operations}
    \label{fig:performance}
    \vspace{-0.3cm}
\end{figure*}
The hardware-based storage experiment is performed with a smartphone and a NFC token. \\
Additionally, we use a laptop for the software-based storage experiment. We repeat the experiment 15 times. \\\\Each iteration takes place as follows:
\begin{enumerate}
\item First, two key pairs are generated: \textit{kp1} that is considered as valid, and \textit{kp2} which issues invalid 
signatures. It means a signature is valid only if it was issued by \textit{kp1}. 
\item After the key generation, \textit{kp1} issues 1000 signatures, and \textit{kp2} issues 200 signatures. 
\item Finally, the issued signatures are verified using \textit{kp1}.
\end{enumerate}
Figure \ref{fig:performance} shows the performance of RSA and ECDSA executing cryptographic operations: generation, 
signing, and verification. ECDSA has a better performance regarding key generation and issuing signatures. RSA, however, is more efficient for verifying signatures. 
Note that the hardware-based approach offers a poor performance for signing operations, since the processor embedded in the NFC token is relatively slow. 
Notwithstanding, this result depends directly on the capabilities in terms of cryptographic operations supported by the token. 
For example, if a token has a dedicated cryptographic co-processor for such operations, it will be faster compared with another one who performs cryptographic operations on the regular micro-processor. 
\section{Discussion and Conclusion}
\label{sec:conclusion}
This paper presented \gls{sol}, a practical framework to bootstrap security for device-to-device settings. \gls{sol} is designed and implemented as an Android service. 
\gls{sol} uses asymmetric cryptographic algorithms: RSA and ECDSA. It also implements a simplified version of the \gls{wot} paradigm. 
It features a Trust Management Layer and a Key Management Layer. The former manages all operations and methods related with the trust relations: bootstrapping, maintaining and synchronization. It deals also with the OoB key verification. The latter performs all operations concerning to the underlying keys: initialization, generation and management. It supports both software and hardware-based key storage solutions.\\
Furthermore, third party apps which utilize our library can benefit from our framework by offering an authenticated and secure communication. To this end, they can create and register the application-specific sub-keys in our service.
Finally, the implementation of a proof-of-concept demons-trates the feasibility of our solution on real devices.
Simulation results confirm the trade-off between trust transitivity and synchronization overhead: transitive trust facilitates a much faster coverage, while direct trust is conserving bandwidth, but limits trust coverage. We make our source code publicly available \cite{solgit}.\\
We foresee a number of performance improvements.
Introdu-cing the concept of a timeout interval together with a register to remember users and timestamp of previous encounters can help saving bandwidth. Then, a new synchronization with a specific user is only allowed after the timeout has expired.  
Furthermore, we can also employ bloom filters for checking already known or trusted users. Thus, it constrains the query part from the synchronization phase. \\
Finally, even tough our scheme builds on direct physical interaction of users and the trust level of transitive relations can be configured. I.e. there is a certain control on how far information generated by malicious devices will spread into the network. 
There are still open issues that are only partially covered or not covered by our solution, for example, an important aspect remains the implementation of methods focused on the revocation of compromised keys in decentralized networks. 
Indeed, key revocation plays an importante role for keeping security in a network as a compromised key can affect the trust of the system partially or totally.
However, as summarized in \cite{ge2016survey}, there is no one-for-all key revocation scheme which deals with all the security issues and requirements of such networks. 
For example, many of them assume the existence of a central authority or a prior known of the network topology, which is not always possible especially in mobile real-world systems, where devices can join and leave the network arbitrarily.
Several schemes rely on the information provided by the devices in the network in order to deal with misbehaving users. However, this can also be used by malicious users to affect the network and to disable legitimate devices.



\section*{Acknowledgment}
This work has been funded by the German Federal Ministry for Education and Research (BMBF) within the SMARTER project.



%
\bibliographystyle{IEEEtran}
\bibliography{References}

\end{document}